\documentclass[12pt]{article}
\usepackage{epsfig,graphicx}

\title{Lepton  decay constants  of light mesons}
\author{  Yu.A.Simonov,\\ Institute of Theoretical and Experimental
Physics\\ 117218, Moscow, B.Cheremushkinskaya 25, Russia}
\date{}

\newcommand{\be}{\begin{equation}}
\newcommand{\ee}{\end{equation}}

\def\fun#1#2{\lower3.6pt\vbox{\baselineskip0pt\lineskip.9pt
\ialign{$\mathsurround=0pt#1\hfil ##\hfil$\crcr#2\crcr\sim\crcr}}}

\newcommand{{\SD}}{\rm SD}

\newcommand{\vex}{\mbox{\boldmath${\rm x}$}}
\newcommand{\vey}{\mbox{\boldmath${\rm y}$}}
\newcommand{\ver}{\mbox{\boldmath${\rm r}$}}
\newcommand{\vesig}{\mbox{\boldmath${\rm \sigma}$}}

\newcommand{\veP}{\mbox{\boldmath${\rm P}$}}
\newcommand{\vep}{\mbox{\boldmath${\rm p}$}}

\newcommand{\vez}{\mbox{\boldmath${\rm z}$}}

\newcommand{\veta}{\mbox{\boldmath${\rm \eta}$}}
\newcommand{\veB}{\mbox{\boldmath${\rm B}$}}
\newcommand{\veH}{\mbox{\boldmath${\rm H}$}}
\newcommand{\veE}{\mbox{\boldmath${\rm E}$}}

\newcommand{\vegam}{\mbox{\boldmath${\rm \gamma}$}}

\newcommand{{\Mc}}{\mathcal{M}}
\newcommand{\llan}{\langle\langle}
\newcommand{\rran}{\rangle\rangle}
\newcommand{\lan}{\langle}
\newcommand{\ran}{\rangle}

\begin{document}

\maketitle
\begin{abstract}
 A theory  of lepton decay constants  based on the  path-integral formalism is given  for chiral and
vector mesons. Decay constants of the
  pseudoscalar and vector mesons are calculated  and compared to other existing
 results.

\end{abstract}

\section{Introduction}

The decay constants $f_n$ in many cases may be directly measured in experiment
and are important characteristics of mesons, where different theoretical
approaches may be compared and their accuracy be estimated. The  $f_n$  of
light mesons have been studied in potential models \cite {1,2,3,4,5,6,7,8,9} in
the QCD sum rule method \cite { 9a,9b,9c, 10,11,12,13,14}  in chiral
perturbation theory \cite {15,16}, as well as in lattice simulations \cite
{17,18,19,20,21,22,23,24} and in experiment \cite{25,26,27,28}.

 The
important role of $f_n$ in theory and experiment is well illustrated by $f_\pi$
- the pion decay constant - which is the basic element and the natural scale of
the chiral perturbation theory \cite{29}. In the latter case $f_\pi$ is taken
from experiment, and the computation of $f_\pi$ from the first principles is a
serious challenge for the theory. On the lattice side a  reasonable accuracy in
computing $f_n$ was achieved recently \cite{17,18,30}, analytic methods include
earlier attempts in the instanton vacuum \cite{31} and within the Field
Correlator Method (FCM) \cite{32,33,34}.

The present article is devoted to the systematic derivation of meson Green's
functions and decay constants $f_n$ for channels with arbitrary quantum
numbers.

This paper is an update and extension of the earlier papers \cite{35}, devoted
to the heavy-light pseudoscalar and light vector  mesons and \cite{36}, devoted
to heavy-light mesons.  Those papers appeared before the systematic formulation
of FCM and in particular of the path-integral  Hamiltonian based on FCM
\cite{37}, therefore some steps in \cite{35,36} required a corrections. In
particular, the rigorous derivation of  the path-integral expression for the
Green's function in \cite{37} has allowed to obtain the improved expressions
for decay constants, which we exploit in what follows. Moreover Chiral Symmetry
Breaking (CSB) was not incorporated in \cite{35,36}. In the  present paper we
present the consistent and general treatment of the meson Green's functions and
its spectral properties also
 for pseudoscalars accounting for CSB in the lowest states ($\pi, K$). The main problem
which one encounters, when addressing the spectral properties in QCD, is the
necessity of the quantitative nonperturbative methods, which сап describe the
main dynamical phenomena: confinement and CSB.

 The familiar (relativistic)
potential model lacks the latter, while other models like instanton vacuum
model,   lacks the former QCD phenomenon.

In what follows we are using the Field  Correlator Method, which was introduced
in \cite{32,33,34}  has acquired  the full form in \cite{37} as a main tool to
study and explain confinement. With respect to the QCD spectrum one derives in
the FCM the effective Hamiltonian, which comprises confinement and relativistic
effects, and contains only universal quantities: string tension $\sigma$,
strong coupling $\alpha_s$ and current quark
     masses $m_q$.  The simple local form of the Hamiltonian which will be called
the   path-integral Hamiltonian    (PIH) occurs for objects with temporal
scales larger than the gluon correlation length $\lambda\approx  0.2 $ fm, i.e.
it is applicable for all QCD bound states except possibly toponium.

 Explicit calculations of masses and wave functions using PIH
have been  done recently for light mesons \cite{38}, heavy quarkonia \cite{39}
and heavy-light mesons \cite{40}, and demonstrate good agreement with
experimental masses.

 In the present paper  we devote a special attention to the chiral mesons and
 trat the chiral symmetry breaking (CSB) phenomenon within  the formalism of
 \cite{41}, where CSB is the consequence of the confinement and the necessary
 relations can be  derived from the basic parameters of QCD: string tension
 $\sigma$ and vacuum correlation length $\lambda$, so that a fundamental
 quantity entering $f_\pi,f_k$ and $m_\pi, m_k$  is $M(0) \approx \sigma
 \lambda$ \cite{41}.

 The paper is organized as follows: in section 2
the general path integral  form of the meson Green's function is presented,
while in Appendices 1-5 the details of derivation are given. In section 3 the
obtained expression for $f_n$ is analyzed. In section 4 light pseudoscalar
mesons are considered. The vector meson decay constants are studied in section
5.  Section 6 contains summary and  concluding remarks.

\section{ The meson Green's function in   the path integral formalism }

We start with the one-body Green's function.

The path-integral representation for $S_i$ is \cite{33,34} \be S_i (x,y) = (m_i
- \hat D^{(i)})\int^\infty_0 ds_i (Dz)_{xy}e^{-K_i}\Phi_\sigma^{(i)}
(x,y)\equiv (m_i-\hat D^{(i)}) G_i (x,y), \label{1}\ee where

\be K_i = m_i^2 s_i + \frac14\int^{s_i}_0 d\tau_i
\left(\frac{dz_\mu^{(i)}}{d\tau_i}\right)^2,\label{2}\ee

$$\Phi^{(i)}_\sigma
(x,y) =P_AP_F \exp \left( ig \int^x_y A_\mu dz_\mu^{(i)}\right)\times$$

 \be
\times \exp \left( \int^{s_i}_0 d\tau_i \sigma_{\mu\nu}
(gF_{\mu\nu}\right).\label{3}\ee Here $F_{\mu\nu} $is  a  gluon  field tensors,
$P_A, P_F$ are ordering operators,  $\sigma_{\mu\nu} = \frac{1}{4i}
(\gamma_\mu\gamma_\nu-\gamma_\nu\gamma_\mu)$. Eqs. (\ref{1}-\ref{3}) hold for
the quark, $i=1$, while  for the antiquark one should reverse the signs of $g$.
In explicit form one writes \be \sigma_{\mu\nu} F_{\mu\nu} = \left(
\begin{array}{ll} \vesig\veH&\vesig \veE\\\vesig\veE& \vesig
\veH\end{array}\right) .\label{4}\ee

The two-body $q_1 \bar q_2$ Green's function can be written as \cite{17, 21}

\be  G_{q_1\bar q_2} (x,y) = \int^\infty_0 ds_1 \int^\infty_0 ds_2
(Dz^{(1)})_{xy} (Dz^{(2)})_{xy} \lan  \hat TW_\sigma (A)\ran_A
e^{-K_1-K_2},\label{5}\ee where \be \hat T = tr (\Gamma_1 (m_1 -\hat D_1)
\Gamma_2 (m_2 -\hat D_2)),\label{6}\ee ``tr'' is the trace over Dirac and color
indices acting on all terms.  Here $\lan W_\sigma (A)\ran$ is the closed Wilson
loop with the spin insertions and one should have in mind, that color spin
insertions in general do not commute, which should be taken into account when
computing spin-dependent part of interaction, see \cite{42}, in (\ref{5}) this
fact was disregarded.
 \be
W_\sigma (A) = P_a P_F \exp \left[ ig \oint A_\mu dz_\mu + g \int^{s_1}_0
\sigma^{(1)}_{\mu\nu} F_{\mu\nu} d\tau_1 - g \int^{s_2}_0 \sigma^{(2)}_{\mu\nu}
F_{\mu\nu} d \tau_2\right].\label{7}\ee

As a result of the correlator averaging \cite{33,34}, and neglecting the
spin-dependent terms, one obtains

 \be \llan W \rran = Z_W \exp (-\int^T_0 [V_0(r(t_E))] )
 dt_E),\label{8}\ee
 where $r(t_E) = |\vez_1 (t_E) - \vez_2(t_E)|$, and

\be V_0 (r) = V_{conf} (r) + V_{OGE} (r), \label{8a}\ee \be V_{conf} (r) = 2r
\int^r_0 d\lambda \int^\infty_0  d\nu D(\lambda, \nu) \to \sigma r, (r\to
\infty),\label{9}\ee

\be \sigma=2\int_0^\infty d\nu\int_0^\infty d\lambda D(\nu,\lambda), \label{10}
\ee

 \be V_{OGE} = \int^r_0 \lambda d\lambda \int^\infty_0
d\nu D_1^{ pert} (\lambda, \nu) =- \frac43 \frac{\alpha_s}{r}\label{11}\ee

At this  point it is useful to introduce  as in \cite{37} the virtual quark
(antiquark)  energies $\omega_1 (\omega_2)$ instead of proper times: $s_i =
\frac{T}{2 \omega_i}$ , where $T$ is the  Euclidean time interval, $T=x_4-y_4$.
As a result one obtains

\be \left( \frac{1}{(m^2_1 - \hat D^2_1) (m^2_2 - \hat D^2_2)}\right)_{xy} =
\frac{T}{8\pi} \int^\infty_0 \frac{d\omega_1}{\omega_1^{3/2}} \int^\infty_0
\frac{d\omega_2}{\omega_2^{3/2}} (D^3 z_1)_{\vex\vey} (D^3 z_2)_{\vex\vey} e^{-
A(\omega_1, \omega_2, \vez_1, \vez_2)},\label{12}\ee where $A\equiv K_1
(\omega_1)  + K_2 (\omega_2) + \int V_0 (r(t_E)) dt_E$, and $$ K_i(\omega_i) =
\frac{m_i^2 +\omega_i^2}{2\omega_i} T + \int^T_0 dt_E \frac{\omega_i}{2} \left(
\frac{d\vez^{(i)}}{dt_E}\right)^2$$ We can also introduce here the two-body 3d
Hamiltonian $H(\omega_1, \omega_2, \vep_1, \vep_2$) and rewrite (\ref{12}) as

\be \left( \frac{1}{(m^2_1 - \hat D^2_1) (m^2_2 - \hat D^2_2)}\right)_{xy} =
\frac{T}{8\pi} \int^\infty_0 \frac{d\omega_1}{\omega_1^{3/2}} \int^\infty_0
\frac{d\omega_2}{\omega_2^{3/2}} \lan \vex |e^{- H(\omega_1, \omega_2, \vep_1,
\vep_2)T}|_{\vey}\ran.\label{13}\ee where $H$ is obtained in a standard way
from the action $A(\omega_1,\omega_2, \vez_1, \vez_2)$ (we omit all e.m. fields
except for external magnetic fields $\veB$)

\be H= \sum^2_{i=1} \frac{(\vep^{(i)})^2 + m^2_i + \omega^2_i}{2\omega_i}
+V_{0} (r) + V_{ss}  + \Delta M_{SE} \label{14}\ee and $V_0$ is given in
(\ref{9}). The spin-dependent part of $H, V_{ss}$ and $V_{LS}$ are obtained
perturbatively from $\sigma_{\mu\nu} F_{\mu\nu}$ terms in (\ref{27}), and is
calculated also  in the presence of m.f. in \cite{42}. It is considered as a
perturbative correction and is a relativistic generalization of the standard
hyperfine interaction,
$$V_{ss} ( r) = \frac{1}{4\omega_1\omega_2} \int \lan \sigma_{\mu\nu}^{(1)}
F_{\mu\nu}(x) \sigma_{\rho\lambda}^{(2)} F_{\rho\lambda}(y)\ran d(x_4-y_4).$$
Its explicit form is given in \cite{42}. Finally, the correction $\frac{\lan
\sigma^{(i)} F(x) \sigma^{(i)} F(y)\ran}{4\omega_1\omega_2}$, where $i$ refers
to the same quark (antiquark) yields the spin-independent self-energy
correction $\Delta M_{SE}$ which
 was calculated
earlier \cite{43}  and for zero mass quarks and no m.f. is  \be \Delta M_{SE}
=- \frac{3\sigma}{2\pi\omega_1}- \frac{3\sigma}{2\pi\omega_2}.\label{14}\ee For
the case of nonzero m.f. the resulting $\Delta M_{SE}$ is given in \cite{42}.
We can now write the total Green's function of $q_1\bar q_2$ system, denoting
by $Y$ the product of projection operators $Y= \Gamma (m_1 - \hat D_1) \Gamma
(m_2 - \hat D_2)$, \be m_1 - \hat D_1 = m_1 - i\hat p_1=m_1 +\omega_1 \gamma_4
- i \vep\vegam ,~~ m_2 - \hat D_2= m_2 - \omega_2 \gamma_4 - i\vep\vegam,
\label{16}\ee where $\vep$ is the quark 3 momentum in the c.m. system.

As a result one has$$\int d^3 (\vex-\vey) G(x,y) = \int d^3 (\vex-\vey)
 tr \left( \frac{4Y_\Gamma}{(m^2_1 - \hat D^2_1) (m^2_2 - \hat
D^2_2)}\right)_{xy}=$$ \be = \frac{T}{ 2\pi} \int^\infty_0
\frac{d\omega_1}{\omega_1^{3/2}} \int^\infty_0 \frac{d\omega_2}{\omega_2^{3/2}}
 \lan Y_\Gamma\ran  \lan \vex |e^{- H(\omega_1, \omega_2, \vep_1,
\vep_2)T}|_{\vey}\ran,\label{17}\ee

We have used  in (\ref{17}) the relations  $ 4\lan Y\ran = tr \lan \Gamma (m_1
- i\hat p_1) \Gamma(m_2 - i\hat p_2))$, and neglect spin dependent terms in
$H$; we have taken into account, that $D_\mu$ acting on Wilson line, i.e.
$D_\mu$ $\exp (ig \int^x A_\mu dz_\mu)\Lambda$ yields $\exp (ig \int^x A_\mu
dz_\mu)
\partial_\mu \Lambda$. The c.m. projection of the Green's function yields  \be
\int d^3 (\vex-\vey) \lan \vex| e^{-H(\omega_1, \omega_2, \vep_1,
\vep_2)T}|_{\vey}\ran = \sum_n \varphi^2_n (0) e^{- M_n (\omega_1, \omega_2)
T}.\label{18}\ee   Here $M_n (\omega_1, \omega_2)$ is the eigenvalue of
$H(\omega_1, \omega_2, \vep_1, \vep_2)$ in the c.m. system, where $\veP =
\vep_1 + \vep_2 =0; ~~ \vep_1 =\vep=-\vep_2$.

The integrals over $d\omega_1, d\omega_2$   for $T\to \infty$  can be performed
by the stationary point method, namely  one has $$ \int G(x,y) d^3 (\vex-\vey)
= \frac{T}{2\pi} \int^\infty_0 \frac{d\omega_1}{\omega_1^{3/2}}\int^\infty_0
\frac{d\omega_2}{\omega_2^{3/2}} \sum_ne^{-M_n (\omega_1,
\omega_2)T}\varphi_n^2(0) \lan Y \ran $$

\be = \sum_n \frac{  e^{-M_n (\omega_1^{(0)}, \omega_2^{(0)}
)T}\varphi_n^2(0)\lan Y\ran }{  \omega_1^{(0)} \omega_2^{(0)} \sqrt{
(\omega_1^{(0)} M^{"}_n(1))(\omega_2^{(0)} M^{"}_n(2))}}, \label{19}\ee where
  \be \left.\frac{\partial M_n (\omega_1, \omega_2)}{\partial \omega_i}
\right|_{\omega_i = \omega_i^{(0)}}=0, ~~ \left. M_n^{"} (i) =\frac{\partial
M_n (\omega_1, \omega_2)}{\partial \omega_i^2} \right|_{\omega_i =
\omega_i^{(0)}},\label{20}\ee
and we have neglected the mixed terms $\frac{\partial^2M_n}{\partial \omega_1
\partial \omega_2}$ for simplicity, however should keep them in concrete
calculations: see exact result in Appendix 1. Comparing the results (\ref{18}),
(\ref{19}) with the definitions of quark decay constants $f^n_\Gamma$,
\begin{eqnarray}
 \int G_\Gamma(x) d^3 \vex & = &
 \sum_n \int d^3 \vex \lan 0 | j_\Gamma| n\ran \lan n |j_\Gamma|0\ran
 e^{i\veP    \vex -M_nT}\frac{d^3\veP  }{2M_n(2\pi)^3}
\nonumber \\
 & = & \sum_n \varepsilon_\Gamma \otimes \varepsilon_\Gamma
 \frac{(M_n f_\Gamma^n)^2}{2M_{n}} e^{-M_nT},
\label{21}
\end{eqnarray}
where for   $\Gamma =\gamma_\mu, ~ \gamma_\mu\gamma_5$ \be
 \sum_{k=1,2,3} \varepsilon_\mu^{(k)} (q) \varepsilon_\nu^{(k)} (q) =
 \delta_{\mu\nu}-\frac{q_\mu q_\nu}{q^2},
\label{22} \ee
and $\varepsilon_\Gamma =1$ for $\Gamma=1,\gamma_5$, one obtains the expression
for $f^n_\Gamma$ (to lowest order in $V_{ss}$)

\be (f_\Gamma^n)^2=\frac{ N_c \lan Y_\Gamma\ran | \varphi_n (0)|^2}{
\omega_1^{(0)} \omega_2^{(0)} M_n \xi_n}, ~~ \xi _n \equiv
\sqrt{(\omega_1^{(0)}M^{"}_n(1))(\omega_2^{(0)} M^{"}_n(2))},
 \label{24} \ee

  This expression coincides with the previously derived in \cite{30,31}, when
  $\xi_n =1/2$. In what follows we show, that $\xi_n$ is close to that value,
  but different for light, heavy-light and heavy-heavy mesons.

\section{Analysis of the obtained expressions }

First calculate $M_n, \varphi_n$ from the equation \be \bar H \varphi_{n} =
M_{n}{\varphi_n},\label{25}\ee
 treating $V_{LS}, V_{SS}$  and $V_{SE}$ in (9) as perturbation, $\bar H = \bar
 H^{(0)}+V_{LS}+ V_{SS}+\Delta M_{SE} = M^{(0)}_n+\Delta M_n$.  One сап simplify the procedure
 introducing the relative coordinate in the c.m. system,
 $\veta = \ver_1-\ver_2$, $\vep= \frac{ \partial}{i\partial \veta}$, so that without magnetic field
\be H_0 =\frac{\vep^2}{2\tilde \omega} +\sum \frac{m_1^2+\omega^2_i}{2\omega_i}
+ V_0 (r) +\Delta M_{SE}, \label{26}\ee
$$H=H_0 +V_{SS} +V_{LS}.$$
(Note, that this is not nonrelativistic expansion!) and finding stationary
values of   $M_n^{(0)}, ~~ H_0 \varphi_n = M_n^{(0)} \varphi_n$ with respect to
$\omega_i, \omega_i = \omega_i^{(0)}$ ) from the equation

\be\left. \frac{\partial M_n^{(0)}(\omega)}{\partial \omega_i} \right|
_{\omega_i = \omega_i^{(0)}} = 0 .\label{27}\ee

This is the basic approach in the string Hamiltonian formalism \cite{44} and it
was checked that the accuracy
 of the replacement $\bar \omega_i=\omega_i^{(0)}$ for lowest states is around 5\% \cite{44}.
  However the values $\varphi_n(0)$ are more sensitive to the replacement (23), and one should use original
   Hamiltonian (15) to calculate $\varphi_n(0)$ \cite{35,36}, see Table  4  of \cite{35}  for comparisons.

It is essential, that we are using $V_{SS}+ V_{LS}$ as perturbation terms to
compute the final hadron masses and hence $M_n$ in different parts of
(\ref{24}) is inserted as computed from $H_0$ (\ref{26}), not containing
$V_{SS} + V_{LS}$.

 ii) As it was argued above, the factor $Y_\Gamma$ can
сап be computed in terms of momenta of quark and antiquark, or in the
c.m.system in terms of relative momentum $\vep$, with the result.


\be \bar Y_V  = m_1m_2+ \bar \omega_1\bar\omega_2 +\frac13 \vep^2
,\label{28}\ee

\be
 Y_{ S}  = -m_1m_2+ \bar \omega_1\bar\omega_2 |+ \vep^2
,\label{29}\ee

\be \bar Y_{A_i}  = - m_1m_2+ \bar \omega_1\bar\omega_2 +\frac{\vep^2}{3}
,\label{30}\ee

\be  Y_{A_4}  = m_1m_2+ \bar \omega_1\bar\omega_2 - {\vep^2}  ,\label{31}\ee

\be  Y_{P}  =( m_1m_2)+ \bar \omega_1\bar\omega_2 - {\vep^2}  ,\label{32}\ee


 Here
  we used   notations:
  \be \hat
Y_V=\sum_i\frac13 tr [(m_1-\hat D_1) \gamma_i (m_2-\hat
D_2)\gamma_i],\label{33}\ee \be \hat Y_{A_i}=-\frac13\sum_i tr [(m_1-\hat D_1)
\gamma_i\gamma_5 (m_2-\hat D_2)\gamma_i\gamma_5],\label{34}\ee
 \be
\hat Y_{A_4}= -[tr (m_1-\hat D_1) \gamma_4\gamma_5 (m_2-\hat
D_2)\gamma_4\gamma_5]. \label{35}\ee

In case of  the pseudoscalar  channel  in (\ref{31}), (\ref{32})  in the chiral
limit\\ $m_1,m_2\to 0$ there appear additional mass terms due to CSB, which are
computed through field correlators and are given in \cite{41}. As it is shown
in Appendix 2, the proper account of CSB leads to the fact, that Eq. (\ref{24})
for $f^2_P$ retains its form for $\pi, K$ mesons, but the expression for
  chiral mesons e.g. $Y_{A_4} $ should be replaced by а more general one, \be Y_{A_4}^{(chiral)} =
(m_1 + M_1 (0))(m_2+M_2(0)) + \bar \omega_1 \bar \omega_2-\vep^2\label{36}\ee

In the chiral limit $m_1=m_2=0$  it was found in \cite{41} that $ M_1 (0) =
M_2(0) \cong 0.15 $ GeV and it was computed through the field correlators,
$M(0) = \sigma\lambda$, where $\lambda$ is the vacuum correlation length,
$\lambda \approx 1$ GeV$^{-1}$ \cite{45}.

 In the nonrelativistic   limit, $m_i\gg
\sqrt{\sigma}$, one can easily find  that $ \bar \omega_i \approx m_i$,  while
$\lan \vep^2\ran \sim O(\sigma)$ , and therefore one has  \be \lan Y_V\ran_{NR}
\approx 2m_1m_2 +0(\sigma),~~\lan Y_S\ran_{NR}\approx 0(\sigma),\label{37}\ee
\be \lan Y_{A_i}\ran_{NR}\approx 0(\sigma);~~ \lan Y_{A_4}\ran_{NR}= 2m_1m_2
+0(\sigma),\label{38}\ee \be \lan Y_P\ran_{NR} = 2m_1m_2 +0(\sigma).
\label{39}\ee Therefore  in  the nonrelativistic limit $m_1\gg \sqrt{\sigma}$,
 $m_2\gg
\sqrt{\sigma}$, for $f_{n}$
 for
   V and P channels one  obtains
\be  f^2_n (NR) =\frac{4N_c}{M_n} |\varphi_n (0)|^2 \label{40} \ee as was found
earlier \cite{1}.

  As a final step one needs   to compute the radiative corrections to
$f_n $, which  come from the short-distance (large momentum) perturbative gluon
contributions. Neglecting interference terms they can be written
 as  in \cite{20,29},
  \be \lan W_\sigma\ran
= \lan W_{OGE} \ran \lan W_{nonpert}\ran\label{41}\ee and
 \be \lan W_{OGE}\ran
= Z_m \exp \left( -\frac{4}{3\pi} \int \int \frac{dz_4 dz'_4 \alpha_s
(z-z')}{(z-z')^2} \right), \label{42}\ee where  $Z_m$  is a regularization
factor. After separating  the Coulomb interaction in  $\hat H$ in this  way,
one gets the correction to $\lan W_\sigma\ran$, and  $f^2_\Gamma$ can be
 written as
 \be
 f^2_\Gamma\to \xi_\Gamma  f^2_\Gamma, ~~ \xi_\Gamma  =1+c_\Gamma\alpha_s + O(\alpha^2_s).\label{43}\ee
but this correction is small and will be  neglected  below.

 Another important contribution from perturbative gluon exchanges
 (GE)
 is the account of the  running coupling constant in  (40) which is   especially important for    $\varphi_n(0)$ in the $S$-wave
 channels. Introducing the  asymptotic freedom factor $P_{AF}$ (we follow notations from
 \cite{35}
 )
\be
 \rho_{AF} =\left | \frac{\varphi_n^{(AF)}(0)}{\varphi_n (0)}
 \right|^2,\label{44}\ee
one should multiply   $f^2_n$  with  this factor  and finally gets
\be
  f^2_\Gamma = \tilde f^2_\Gamma ~~\xi_\Gamma\rho_{AF}.\label{45} \ee

   We conclude this section with the discussion of input parameters
in the
 approach described above. The set of parameters
includes    $m_i,  \alpha_s$ and $\sigma$ in the first approximation   and
$m_i, \alpha_s (r),\sigma$, when asymptotic freedom is taken into account.

 Here $m_i$  are
pole masses which are connected to the Lagrangian (current) masses in
$\overline{MS}$ scheme as  (see \cite{25} for а review and \cite{36} for
additional references).

\be m_i =\bar m_{\overline{MS}} ( \bar m_{\overline{MS}}) \left\{ 1+ \frac43
\frac{\alpha_s (\bar m_{\overline{MS}})}{\pi} +\eta_2
\left(\frac{\alpha_s}{\pi}\right)^2 + O({\alpha_s})^3\right\}. \label{46}\ee

\section{ Light pseudoscalar mesons and current correlators in PS
channels}

The formalism of the previous section is of general character and сап be
applied to any channel $\Gamma$. However to save space we shall consider below
only PS and  vector  mesons. Pseudosclalar mesons appear both in A$_4$ and P
channels. Their connection to the A channel is given by  the chiral anomaly
term:

\be \lan 0| j_\mu^{(A)}|q,n \ran = i f_P (n) q_\mu.\label{47}\ee

Exploiting this definition one obtains the same expression as before for
$f_P(n)$, considering the c.m. system $\veP = 0$, namely \be f_P^2(n) =\frac{
N_c \lan Y^{(chiral)}_{A_4} \ran | \varphi_n (0)|^2}{\bar \omega_1 \bar
\omega_2 M_n\xi_n}.\label{48}\ee

One should have in mind however that our  formalism above in this paper till
now did not take into account Chiral Symmetry Breaking (CSB) and therefore
cannot be applied to the Nambu-Goldstone mesons $\pi, K, \eta$. For the latter
one should use the technic suggested and exploited in \cite{41}, where $f_\pi$
was computed through the masses $m_n, \varphi_n (0)$  as in (\ref{24}) but in
addition there appears an effective mass parameter $M(0)$,  see Appendix 2. The
resulting equation for $\lan Y_{A_4}\ran$   Eq. (\ref{36}) can be written  in
the  limit $m_i \to 0$ as

\be \lan Y_{A_4}\ran  =  M^2 (0) + \lan \sqrt{\vep^2 + m^2_1}\ran \lan
\sqrt{\vep^2 + m^2_\bot}\ran-\lan \vep^2\ran \to M^2(0).\label{49}\ee

In the chiral limit, $m_1 =m_2 =0$ , and taking into account that $\bar
\omega_1 = \bar \omega_2 = \bar \omega$,  and $\lan \vep^2\ran  = \bar
\omega^2$ ,  one has

\be f_P^2(n) =\frac{ N_c  M^2 (0) | \varphi_n (0)|^2}{ \bar \omega^2 \bar
  M_n\xi_n }.\label{50}\ee

For $\varphi_n(0)$  and $M_n$ one takes neglecting hyperfine interaction the
same values, as for $\rho$-meson, i.e. $ M_n = \bar M (n=0) =0.65$  GeV,
$|\varphi_n (0) |^2= \frac{0.109 GeV^3}{4\pi}, ~~ \bar \omega = 0.352 $ GeV.

 Taking now $\xi_n^{-1}= 2.45$ from A(12) and $M(0)=0.15$ GeV one obtains $f^2_\pi = 0.01782$ GeV$^2$, $f_\pi =133$ MeV.
  This should be compared with the experimental value, which in the normalization of Eq.
  (\ref{50})
 is equal to $f^{ex}_\pi\cong  \sqrt{2} \cdot 0.093$ GeV $= 0.131 $ GeV. One сап see that
 agreement within 2\%.

One should stress, that in absence of CSB, when $M(0) \equiv 0, $ and $\lan
Y_{A_4}\ran \to m_1m_2 \to 0$,  also $f_\pi$ vanishes, implying that $f_\pi$
plays the role of the CSB order parameter (together with $\lan  q\bar q\ran $,
which is also proportional to $M(0)$).

We now turn to the case of $K $ meson. Doing calculations in the same  way as
 for pion above,  and taking $m_s = 0.15$
GeV,  $\sigma = 0.18$ GeV$^2$, one has for $K$- meson;

$$ \omega_u (K) =   0.36 {\rm GeV},~~ \omega_s(K) = 0.39 {\rm GeV,}~~  M_K^{(0)} = 0.84 ~{\rm GeV}. $$

The latter number is obtained without Coulomb and hyperfine interaction, which
shift $ M_k$ by $\Delta m_{Coul}  = 0.05$ GeV and $\Delta  m_{Hf} =0.3$ GeV,
resulting in $ m_K = M_K^{(0)} -\Delta m_{coul} - \Delta m_{Hf} \cong 0.49$
GeV.

Using Appendix 1, Eq. (A2), one obtains $\xi_K^{-1} = 2.29$ so that $f_K^2$ is
\be f^2_K = \frac{2.29 \cdot N_c \lan Y_K\ran \varphi^2_K (0)}{\omega_u\omega_s
\bar M_K^{(0)}}, \label{51}\ee where $\lan Y_K\ran = (M(0) + m_u ) (M(0) + m_s)
+ \omega_u\omega_S - \lan \vep^2\ran = 0.06$  GeV$^2$ and as a result $f_K
=0.165$ GeV.

To compare $f_K$ and $f_\pi$  we write down for  both mesons without  hyperfine
interaction and using (49)

\be \frac{f^2_K}{f^2_\pi} = 1.6,\label{52}\ee

$$ \frac{f_K}{f_\pi} = 1.24$$

 The result (\ref{52}),  $\frac{f_K}{f_\pi}
= 1.24$  is in   agreement with the experimental values \cite{25}

\be f^{(\exp)}_{\pi^+} = 130.7 \pm 0.1\pm 0.36 {\rm ~ MeV},\label{53}\ee

$$ f^{(\exp)}_{K^+} = 159.8 \pm 1.4\pm 0.44{\rm ~ MeV},$$
which yields

\be \frac{f^{(\exp)}_{K^+}}{f^{(\exp)}_{\pi^+}} =1.22\pm 0.02,\label{54}\ee
while lattice  calculation \cite{17} yield for this ratio $1.195 \pm 0.006$.

We now turn to the radial excitations of the chiral mesons. In this case of
high excitations there appear decay channels, which play  a role of
intermediate channels in the meson Green's function. Therefore neglecting these
channels, we shall make only rough upper limit  estimates of decay constants.
To this end we exploit the fact (see Appendix 1) that $\xi_n$ does not depend
on $n$, and $\varphi_n^2 (0)$ can  be estimated to the  lowest order as
$\frac{\sigma \omega_n}{4\pi} \left|\frac{\chi_n (x)^2}{\chi)n (0)}\right|$ and
$\bar M_n \approx 2 \omega_n$ ( with the accuracy of (10$\div$ 15)\%). As a
result one obtains \be f_{\pi n}  ({\rm~ GeV})\cong \frac{0.105}{M_n({\rm ~
GeV})}, ~~n>1\label{55}\ee where $M_n$ are the mass values before the chiral
shift \cite{46}, and as a result one has the values listed in the Table 1.

\begin{table}
\caption{ Decay constants of chiral mesons and its excitations}
\begin{tabular}{|l|l|l|l||l|l|l|}\hline
\multicolumn{4}{|c||}{$\pi(nS)$}& \multicolumn{3}{|c|}{$K(nS)$}\\\hline

n&1&2&3&1&2&3\\\hline

 $f_{\pi n}$(GeV)&
0.138&0.069&0.048&0.165&0.104 &0.085\\\hline $f_{\pi n}/f_{\pi 1}$
&1&0.5&0.35&1&0.63&0.515\\\hline

\hline \hline
\end{tabular}
\end{table}

One can see, that $f_{\pi n}$ and $f_{K n}$ are slowly decreasing for growing
$n$, implying a substantial leptonic decay contribution to the list of decay
modes.

\section{Decay constants of vector mesons}

 We now turn to vector mesons, where we take the same value for $\bar M_n$ and
 $\xi_n$ (\ref{49}), not affected by the $hf$ splitting, which we take
 afterwards as a perturbation. Hence we take $\xi^{-1}_K = 2.29$ for $K^*$
 meson and $\xi^{-1}_\rho=2.45$ for $\rho$ and $\omega$ mesons, while for the
$ \phi$ meson one obtains  $\xi^{-1}_\phi=2.095$, and  $\omega_s=0.424$ GeV,
$\Delta M_{SE} = - 0.282$ GeV and $M_\phi= 1.040$ GeV. ($ M_\phi(\exp) = 1.020$
MeV).

Since vector mesons are connected to the  electromagnetic current, the
corresponding decay constants contain the effective charge $\bar e^2_q (i)
=\frac12, \frac{1}{18}, \frac19$ for $i=\rho, \omega, \phi$ [] and vector decay
constants have the form \be f_n^{(V)} (i) =  \frac{\bar e^2_q (i) N_c
\varphi^2_n (0) \lan Y_V\ran}{\omega_1^{(0)} \omega_2^{(0)} M_n (i) \xi_n (i)},
i=\rho, \omega, \phi,\label{56}\ee while the dielectron width is connected to
$f_n^{(V)}(i)$ as \cite{25} \be \Gamma(V\to e^+e^-) =
\frac{4\pi\alpha^2}{3M_V(i)}(f_n^{(V)} (i))^2 (1-\frac{16}{3\pi}
\alpha_s).\label{57}\ee

Keeping the value of the last factor in (\ref{57}) to be equal 0.32$(\alpha_s
=0.4)$ for $\rho, \omega$ and  $\alpha_i=0.3$ for $\phi$, as a result one
obtains the  values of  $f^V(\exp)$, given in the Table 2. Lattice data
\cite{19} yield $f_\rho = 239 (18)$ in a satisfactory agreement with our result
in Table 2.

\begin{table}
\caption{Decay constants of vector mesons}
\begin{tabular}{|l|l|l|l|}\hline
$V_i$ &$\rho$&$\omega$&  $\phi$\\\hline

$f^V_{n=0} $ ( GeV)& 0.254&0.0846&0.096\\\hline $f^V (\exp)$
(GeV)&0.255&0.0756&0.107\\\hline $\Gamma_{ee} (\exp)$ (keV)&$7.04\pm 0.06$&
$0.60\pm0.02$& $1.26\pm
0.02$\\

\hline \hline
\end{tabular}
\end{table}

\section{Summary and conclusions}

We have presented the theory of lepton decay constants for light mesons  based
on the path integral formalism. It  essentially exploits the path integral
Hamiltonian (PIH) depending on virtual $q,\bar q$ energies $\omega_1, \omega_2$
and the final form obtains after the stationary point analysis with respect to
$\omega_1, \omega_2$. The same approach has given a large number of observables
(masses and wave functions, Regge trajectories etc.) in good agreement with
experiment both for light and  heavy quarks \cite{38, 39, 40}. The  lepton
decay constants for heavy-light meson have been  calculated in the same method
in an approximate form in \cite{35,36} also in good agreement with lattice and
experimental data.

In this paper we specifically considered light mesons and paid a special
attention to the  chiral mesons and their radial excitations. We have also
exploited an  improved form of the path integral from \cite{37}, which allows
to obtain much better accuracy. Another important ingredient  of the present
paper is  a new treatment of chiral mesons, which  exploits the fundamental
quantity  - the scalar chiral mass parameter   $M(0) \approx \sigma \lambda$
(the corresponding  chiral correlation length is $1/\sigma \lambda$), which as
shown before in \cite{41} and here in Appendix 2, enters additively with the
current quark mass $m_q$ and  disappears at large $m_q$.

The resulting values of $f_\pi$ and $f_K$ are in good agreement with
experimental data, however the only parameters of our theory are $m_q,
\alpha_s$ and string tension $\sigma$. We have also calculated decay constants
of   vector mesons $\rho, \omega, \phi$ and found a satisfactory agreement with
experiment.

The   author is grateful to A.M.Badalian for  useful suggestion and
discussions. Financial support of the RFBR grant 1402-00395 is gratefully
acknowledged.

\vspace{2cm}
 \setcounter{equation}{0}
\renewcommand{\theequation}{A.\arabic{equation}}

\hfill {\it  Appendix  1}

\centerline{\it\large  The correction coefficient $\xi_n$}

 \vspace{1cm}

\setcounter{equation}{0} \def\theequation{A1.\arabic{equation}}

As it was shown in \cite{37} (see also appendix of \cite{37}), $\xi_n$ in
(\ref{24}) is defined as follows: \be \xi_n =
\sqrt{\omega_1^{(0)}\omega_2^{(0)} \Omega_n}, \label{A1}\ee with

 \be \Omega_n =  \left( \alpha \beta - \frac{\gamma^2}{4} \right)
,\label{A2}\ee and \be \alpha \equiv \frac12 \frac{\partial^2 M_n}{\partial
\omega^2_1}, ~~ \beta =\frac12 \frac{\partial^2 M_n}{\partial
\omega^2_2},~~\gamma =  \frac{\partial^2 M_n}{\partial \omega_1\partial
\omega_2},\label{A3}\ee

$M_n$ is defined as \be M_n = \sum_{i=1,2} \frac{\omega^2_i+  m^2_i}{2\omega_i}
+ \varepsilon_n (\tilde \omega^{-1}); ~~ \tilde \omega^{-1}
=\frac{\omega_1+\omega_2 }{ \omega_1\omega_2},\label{A4}\ee and $\varepsilon_n$
is the eigenvalue of the equation \be \left( \frac{\vep^2}{2\tilde \omega} + V
(r) \right) \varphi_n (r) = \varepsilon_n \varphi _n (r),\label{A5}\ee where
$V(r)$ includes confinement $V_c(r)$ and gluon exchange interaction $V_g(r)$,
but not hyperfine interaction, which is taken into account as a first order
correction to the total mass, and hence is not to be present in (\ref{A3}).
Hence finally $\bar M_n$,  entering in $f^2_n$, is the hyperfine  averaged
eigenvalue $M_n$ with the selfenergy term $\Delta_{SE}$ taken into account

\be \bar M_n = \sum \frac{(\omega^{(0)}_i)^2 + m^2_i}{2\omega_i^{(0)}} +
\varepsilon_n ((\tilde \omega^{(0)})^{-1}) + \Delta_{SE}\label{A6}\ee For
$\alpha,\beta, \gamma$ one obtains

\be 2\alpha, 2\beta=\frac{m^2_i}{\omega^3_i} + \frac{2}{\omega^3_i}
\varepsilon'_n + \frac{1}{\omega^4_i} \varepsilon^{''}_n, i=1,2.\label{A7}\ee
\be \gamma= \varepsilon^{''}_n \frac{1}{\omega^2_1 \omega^2_2}\label{A8}\ee
Here $\varepsilon'_n = \frac{\partial \varepsilon_n}{\partial \tilde
\omega^{-1}} , ~~\varepsilon^{''}_n = \frac{\partial^2 \varepsilon_n}{\partial(
\tilde \omega^{-1})^2}$.

For $m_1=m_2=m$ and hence $\omega_1 = \omega_2 \equiv\omega$ one has
$\alpha=\beta$ and $\Omega_n$ is \be \Omega_n =\alpha^2- \frac{\gamma^2}{4} =
\frac{\varepsilon^{''}_n (\varepsilon'_n + \frac{m^2}{2} ) +\omega
(\varepsilon'_n + \frac{m^2}{2} )^2}{4\omega^7}.\label{A9}\ee
 In the  nonrelativistic limit $\varepsilon_n \ll m, \omega \cong m$, one has
 \be \Omega_n \cong \frac{1}{4m^2},  \xi_n =\frac12.\label{A10}\ee

Consider now light quarks and put $m_1 =m_2=0$, and first neglect the OGE
interaction. Then \be \varepsilon_n = 2^{-1/3}(\tilde \omega^{-1})^{1/3}
\sigma^{2/3} a(n), \label{A11}\ee with $a_0 = 2.338$ and other $a(n)$ given in
Table 2 of \cite{44}.

Using (\ref{A7}), (\ref{A8}), one obtains \be\xi_n (\alpha_s = 0 , m_i =0) =
\frac{3}{\sqrt{54}}=0.408; ~~ \xi_n^{-1} = 2.45.\label{A12}\ee

The resulting $f^2_\Gamma$  is \be f^2_{\Gamma, n} = \frac{2.45 N_c \lan
Y_\Gamma\ran \varphi_n^2 (0)}{\omega^2_0 \bar M_n}, \label{A13}\ee where
$\varphi^2_n(0) = \frac{\omega_0\sigma}{4\pi} , \lan Y_V \ran = \frac43
\omega^2_0, \bar M_0 = 4 \omega_0 + \Delta_{SE} \cong 2 \omega_0$.

The inclusion of OGE interaction yields \cite{35,44} \be \varepsilon_n^{(g)} =
\varepsilon_n \frac{a(\lambda, L, n)}{a(n)}=2^{-1/3}(\tilde \omega^{-1})^{1/3}
\sigma^{2/3} a(\lambda, L,n), \label{A14}\ee where $\lambda=
\frac{4\alpha_s}{3} \left( \frac{2\tilde \omega}{\sqrt{\sigma}}\right)^{2/3}$
and $a(0,L,n)=a(L,n).$

The values of $a(\lambda, 0,0), \frac{\partial a}{\partial \lambda} (\lambda,
0,0)$ are given in the Table 4 of \cite{35}, and one has an estimate of
$\frac{\partial^2 a}{\partial \lambda^2} \cong 0.2$ for $\lambda<0.9$.

The most important change due to OGE is in $\varphi^2_n(0)$ which is now for
$L=0$ \be \varphi_n^2 (0, \alpha_s) = \frac{\omega}{4\pi} (\sigma + \frac43
\alpha \lan r^{-2}\ran_n ) = \frac{\sigma \omega}{4\pi} \left | \frac{
\chi_\lambda (0) }{\chi_0(0)}\right|^2\label{A15}\ee

The values of $\varphi^2_n (0, \alpha_s)$ are given in the Table 4 of
\cite{35}, together with $\left | \frac{ \chi_\lambda (0)
}{\chi_0(0)}\right|^2$.

\vspace{2cm}
 \setcounter{equation}{0}
\renewcommand{\theequation}{A.\arabic{equation}}

\hfill {\it  Appendix  2}

\centerline{\it\large   Chiral correction length in the  confining  vacuum}

 \vspace{1cm}

\setcounter{equation}{0} \def\theequation{A2.\arabic{equation}}

It was shown in \cite{41}, that  nonzero field correlator $\lan FF\ran \equiv
 \lan tr (F (u) \phi F (v) \phi\ran$,  generating the kernel $J(x,y) \sim \int
 du \int dv \lan FF \ran$, leads to the   appearance in  the  quark Green's
 function $S(x,y)$ the  nonperturbative nonlocal mass operator $\mathcal{M}
 (x,y)$,
$\mathcal{M}
 (x,y) \sim  JS,$ satisfying equation

\be (\hat \partial + m_q) S (x,y)  + \int \mathcal{M}
 (x,y) S(z,y) d^4 z = \delta^4 (x-y).\label{A2.1}\ee
This general property can be  further analyzed separating in $\mathcal{M} $
scalar-isoscalar part $\mathcal{M}_s$ and  pseudoscalar-isovector pieces $\hat
U$ which can be conveniently written as

\be \mathcal{M} \to \mathcal{M}_S \hat U, ~~ \hat U = \exp (i \gamma_5 \hat
\phi)\label{A2.2}\ee As a consequence the scalar nonlocal mass $\mathcal{M}_S
 (x,y)$ enters into the  quark Greens function $S$ as a scalar piece together
 with the current quark mass $m_q$,
 \be iS (x,y) = (\hat \partial +m_q + M_S)^{-1}_{xy}.\label{A2.3}\ee

 Note, that at large $m_q$, the magnitude of $M_S$ is fast decreasing
 \be  M_S \sim O(1/m_q), ~~ m_q \to  \infty .\label{A2.4}\ee

 In the current correlator $\lan x | \Gamma S_q \Gamma S_{\bar q}|y\ran, M_S$
 enters in the local vertex form $M_S (x,x) = M_S (0)$, and in the framework of
 the chiral approach \cite{41} based on (\ref{A2.1}) and (\ref{A2.2}), one
 derives the relation for $f^2_\pi$ in the chiral limit, $m_q\to 0$,
 \be f^2_\pi = N_c M^2 (0) \sum_{n=0} \frac{|\varphi_n
 (0)|^2}{M^3_n}\label{A2.5}\ee
 where $n$ refers to the radial excited $q\bar q$  states with mass $M_n$, and
 the sum over $n$ is cut off by the factor $\exp (-M_n \lambda), \lambda =0 (1$
 GeV$^{-1})$ is the vacuum correlation length, calculated via gluelump masses
 \cite{45}, $M(0)$ was calculated in the Appendix 4 of the third reference
 \cite{41},
 \be M(0) \cong \frac{2\sigma \lambda}{\sqrt{\pi}} \approx 0.15~ {\rm GeV}.
 \label{A2.6}\ee

\end{document}